# Fabrication and Characterization of Magnetic-Field-Resilient MoRe Superconducting Coplanar Waveguide Resonators

Chang Geun Yu [1], Bongkeon Kim [1] and Yong-Joo Doh[1*]

[1] Department of Physics and Photon Science, Gwangju Institute of Science and Technology, Gwangju, 61005, Korea.

Magnetic-field-resilient superconducting coplanar waveguide (SCPW) resonators are essential for developing integrated quantum circuits of various qubits and quantum memory devices. Molybdenum-Rhenium (MoRe), which is a disordered superconducting alloy forming a highly transparent contact to the graphene and carbon nanotubes (CNTs), would be a promising platform for realizing the field-resilient SCPW resonators combined with graphene- and CNT-based nano-hybrid qubits. We fabricated MoRe SCPW resonators and investigated their microwave transmission characteristics with varying temperature and external magnetic field. Our observations show that the thickness of MoRe film is a critical parameter determining the lower critical field, kinetic inductance, and characteristic impedance of the SCPW resonator, resulting in drastic changes in the quality factor and the resonance frequency. As a result, we obtained a maximum value of $Q_i > 10^4$ in parallel magnetic fields up to $B_\parallel = 0.15$ T for the 27-nm-thick MoRe resonator. Our experimental results suggest that MoRe SCPW resonator would be useful for integrating nano-hybrid spin or gatemon qubits and for developing spin-ensemble quantum memory devices.

**Corresponding Authors**

[*]E-mail: yjdoh@gist.ac.kr

**KEYWORDS:** MoRe, magnetic-field-resilient superconducting coplanar waveguide resonator, nano-hybrid qubits, quality factor, resonance frequency

A superconducting coplanar waveguide (SCPW) is a type of microwave transmission line consisting of a central superconductor sandwiched between two ground planes on a dielectric substrate [1]. By shorting or opening the central superconductor to the ground plane at both or either ends of the SCPW, resonance can be generated for a specific frequency of the propagating microwave. Owing to this property, an SCPW can be used as a microwave resonator maintaining high quality factor [2]. Since the SCPW resonator has the advantage of easy design and fabrication for desired properties, various hybrid devices combined with SCPW resonators have been studied, such as superconducting qubits [3], gatemon qubits [4, 5], spin qubits [6, 7], and spin ensemble memory [8]. In recent years, the resonator operation in an external magnetic field has been required for the optimal performance of the spin-based quantum devices [6-8] and the gatemon qubit [9], but a significant reduction of the internal quality factor, $Q_i$, and a shift of the resonance frequency, $f_c$, of the SCPW resonator in an external magnetic field become a concerning matter [10, 11]. Therefore, research on a magnetic-field-resilient SCPW resonator that maintains high $Q_i$ values under high magnetic field [12] is essential for the implementation of quantum information processing devices.

In recent years, graphene and carbon nanotubes (CNTs) have been used to develop nano-hybrid superconducting qubits [13] or spin qubits [14]. Since Molybdenum-Rhenium (MoRe), which is a disordered superconducting alloy with a relatively high transition temperature, forms a highly transparent interface to the graphene [15] and CNTs [16], MoRe SCPW combined with graphene- or CNT-based qubits would be a promising platform to develop integrated quantum circuits for scalable quantum computing [17]. It has been reported that MoRe SCPW exhibited a maximum value of $Q_i = 7 \times 10^5$ under zero magnetic field [18]. However, the magnetic field resilience of MoRe SCPW resonators, which is essential for developing integrated spin qubits [14, 19] and quantum memory devices [8], has not been studied yet.

It is well known that superconducting thin films in a magnetic field exhibit Abrikosov vortex structures [20]. Since the dissipation due to vortices can result in a substantial reduction of the quality factor of the SCPW resonator [11], it would be crucial to enhance the lower critical field, $B_{c1}$, of the superconducting strips for the complete expulsion of vortices. When the film thickness is smaller than the London penetration depth, $\lambda_L$, the in-plane lower critical field, $B_{c1,\parallel}$, is proportional to $\Phi_0/t^2$, where $\Phi_0$ is the magnetic flux quantum and $t$ is the film thickness [12, 21]. Thus it is expected that thinner superconducting films would be more resilient to a magnetic field, when the magnetic field is applied parallel to the film surface. In this study, we fabricated MoRe SCPW resonators of various film thicknesses ranging from 15 nm to 120 nm and characterized their microwave transmission properties with varying temperature $T$ and the parallel magnetic field $B_\parallel$. As a result, we obtained a maximum value of $Q_i > 10^4$ with $B_\parallel = 150$ mT at $T = 1.8$ K for the 27-nm-thick MoRe resonator. Our experimental results suggest that MoRe SCPW can be used as a useful platform for integrating nano-hybrid superconducting qubits [13], spin qubits [7, 14, 19, 22], and gatemon qubits [9] and for developing a spin-ensemble quantum memory [8, 23].

The MoRe SCPW resonators were fabricated on an undoped Si wafer with a resistivity of about $\rho_{Si} = 10$ kΩ·cm at room temperature, covered by 100 nm of thermally grown $SiO_2$. The MoRe deposition was performed using a pulsed DC sputtering system with a high-purity (99.99 %) $Mo_{0.66}Re_{0.34}$ alloy target. The Ar pressure during deposition was 4 mTorr and the sputtering power was 200 W, resulting a deposition rate of 0.5 nm/s. Figure 1a shows the temperature dependence of the resistivity, $\rho$, of MoRe thin film with different thicknesses near the superconducting transition temperature $T_c$. It is noted that the residual resistivity, $\rho_{res}$, increases and $T_c$ decreases monotonously with decreasing film thickness (see Fig. 1b). The resonators were patterned by photolithography and reactive ion etching in an $SF_6/O_2$ (4 sccm/1 sccm) atmosphere. After the fabrication was completed, the resonator sample was mounted in an Au-coated copper box, in which the central strip line common to all SCPW resonators was connected to bond pads in the sample holder by using ultrasonic bonding of Al wire. Measurements at low temperatures were performed using a closed-cycle refrigerator (Cryogenic Ltd.) with a base temperature of 1.8 K, equipped with a homemade RF probe. The microwave transmission, $S_{21}$, through the sample was measured using a network analyzer (Agilent Inc.). Detailed information is available elsewhere [24].

Figure 2a shows an optical microscope image of our sample containing six quarter-wavelength ($\lambda/4$) resonators coupled to the feed line. Each resonator has a serpentine shape, as shown in Fig. 2b, with an elbow bend shorted to the ground plane (see Fig. 2c) and the opposite end open-circuited (Fig. 2d). The elbow length determines the coupling capacitance between the resonator and the feed line [25]. We also formed the rectangular hole (5.8 × 5.8 μm²) arrays in the ground plane, as shown in Fig. 2d-e, to trap Abrikosov vortices generated by an external magnetic field [26].

The fundamental resonance frequency $f_c$ is dependent on the total length $l$ of the resonator via the relation of $f_c = [(L_g + L_k)C_g]^{-1/2}/4l$, where $L_g$ ($C_g$) means the geometrical inductance (capacitance) and $L_k$ is the kinetic inductance of the SCPW resonator [12]. Using the conformal mapping method [24], we obtained $C_g =$ 168 pF/m and $L_g =$ 424 nH/m from the designed values of the central conductor ($s =$ 12 μm) and slot gap ($w =$ 6.5 μm) widths. Since $l$ ranged from 6.14 mm to 6.50 mm in our design, $f_c$ is expected to vary between 4.58 GHz and 4.85 GHz. Then the characteristic impedance, $Z_0 = \sqrt{(L_g + L_k)/C_g}$ [24, 27], of the resonator is matched to 50 Ω by assuming $L_k \ll L_g$ for now. The experimental parameter values of MoRe SCPW resonators used in this work can be found in Table 1.

After cooling the SCPW sample under zero magnetic field, the $S_{21}$ spectrum measurements were carried out. Figure 3a displays typical $S_{21}$ data as a function of the probe frequency $f$, obtained from 60 nm-thick MoRe SCPW sample at $T =$ 1.8 K. Six resonant dips, identified by each resonator name (R1 − R6), are clearly observed at the fundamental resonant frequencies of the six $\lambda/4$ resonators formed in the sample. For a theoretical fit, we converted $S_{21}$ into transmission coefficient, $S'_{21}$, in Fig. 3b, using the relation $|S'_{21}| = 10^{(|S_{21}|-LL)/20}$, where $LL$ means the line loss in coaxial cables [27]. The complex transmission coefficient $S'_{21}$ is expressed by the following equation [12]:

$$S'_{21} = A\left(1 + \alpha'\frac{f-f_c}{f_c}\right)\left(1 - \frac{Q_l e^{i\theta}/|Q_e|}{1+2iQ_l(f-f_c)/f_c}\right) \quad (1),$$

where $A$ is the amplitude parameter, $\alpha'$ is a background slope parameter, $Q_e = |Q_e|\exp(-i\theta)$ is a complex-valued external quality factor related to a coupling quality factor $Q_c$ via $1/Q_c = \text{Re}(1/Q_e)$, and $Q_l$ is the loaded quality factor satisfying $1/Q_l = 1/Q_i + 1/Q_c$. The solid line in Fig. 3b was obtained from the theoretical fit with $f_c =$ 3.61855 GHz, $Q_i = 1.7 \times 10^4$, and $Q_c = 1.8 \times 10^3$ for R3 resonator with 60 nm thickness. We performed the $S$-parameter fit to other resonance dips of MoRe resonators in this work to extract their respective $f_c$ and $Q_i$ values. We found that R3 resonator had the highest $Q_i$ as compared to other resonators with $t =$ 41, 60, 89, and 120 nm, similarly R4, R5 resonator with $t =$ 15 and 27 nm were the best respectively, so we stuck to the high $Q_i$ resonator for each thickness for the characteristic analyses in this paper.

When temperature increases, the progressive change of the transmission spectrum is shown in Fig. 3c, displaying a gradual shift of $f_c$ to lower frequency and broadening of the resonance dip. The transmission coefficient plot drawn relative to the center of the resonance dip in Fig. 3d depicts more clearly the thermal broadening of the dip. After fitting the $S'_{21}$ data to Eq. (1), we obtained $f_c$ and $Q_i$ values as a function of temperature for the MoRe SCPW resonators with different film thicknesses, as displayed in Fig. 4a-b. Here, $\delta f_{c,T} = f_c(T) - f_c(1.8\ \text{K})$ means the temperature-dependent difference in $f_c$ relative to the value obtained at the base temperature, $f_{c,0}$. We note that the negative shift of $f_c$ and the monotonous decrease of $Q_i$ at higher temperatures are observed in all the MoRe resonator samples studied in this work.

The negative shift of $f_c$ can be understood qualitatively as a consequence of the reduced Cooper-pair density in the superconducting MoRe film at higher temperatures [18], resulting in an increase of the London penetration depth $\lambda_L$ [20]. Since the kinetic inductance is given by $L_k = g\mu_0 \lambda_L^2/st$, where $g$ is a geometric prefactor [28], $L_k$ would increase drastically for thinner films at higher temperatures and become comparable to or even exceed $L_g$. Thus $f_c$ decreases with increasing temperature, resulting negative shift of $\delta f_{c,T}$ in Fig. 4a. The red shift of $f_c$ becomes more sensitive to temperature for thinner films, which is attributed to their lower $T_c$'s (see Table 1). Reduction of $Q_i$ at high temperatures, following asymptotically $T^{-\nu}$ with $\nu = 3.9 \pm 0.4$, is attributed to an increase of quasiparticle density in MoRe film, leading to larger damping in the SCPW resonator [29]. The lower $Q_i$ values in the thinner films are also attributed to their relatively lower $T_c$'s, which enhances the quasiparticle effect at the same temperature.

To be more quantitative, we used the Mattis–Bardeen (M–B) theory [30] to fit our experimental data. According to the M–B theory, the temperature-dependent differences of $\delta f_{c,T}/f_{c,0}$ and $\delta(1/Q_i)$ can be expressed as follows [29, 30]:

$$\frac{\delta f_{c,T}}{f_{c,0}} = \frac{\alpha}{2}\frac{\delta\sigma_2}{\sigma_2} \quad (2)$$

$$\delta\left(\frac{1}{Q_i}\right) = \alpha \frac{\delta\sigma_1}{\sigma_2} \quad (3)$$

Here, $\alpha = L_k/(L_k + L_g)$ is the kinetic inductance fraction in zero-temperature limit and $\sigma_1$ ($\sigma_2$) is the real (imaginary) part of the complex conductivity, $\sigma = \sigma_1 - i\sigma_2$, of the SCPW resonator, which can be expressed by

$$\frac{\sigma_1}{\sigma_n} = \frac{4\Delta(T)}{hf_c} e^{-\frac{\Delta(T)}{k_BT}} \sinh\left(\frac{hf_c}{2k_BT}\right) K_0\left(\frac{hf_c}{2k_BT}\right) \quad (4),$$

$$\frac{\sigma_2}{\sigma_n} = \frac{\pi\Delta(T)}{hf_c}\left[1 - 2e^{-\frac{\Delta(T)}{k_BT}} e^{-\frac{hf_c}{2k_BT}} I_0\left(\frac{hf_c}{2k_BT}\right)\right] \quad (5),$$

where $\sigma_n$ is the normal-state conductivity, $\Delta(T)$ is the superconducting gap energy, $h$ is the Planck constant, $k_B$ is the Boltzmann constant, and $I_0(x)$ and $K_0(x)$ are the modified Bessel function of the first and second kind, respectively [29-31]. The fit curves obtained from the M–B theory using $\alpha$ as a single free parameter are depicted by the solid lines in Fig. 4a-b, which are in good agreement with the experimental data. Since $\sigma_1$ ($\sigma_2$) is dependent on the quasiparticle (Cooper pair) density, the negative shift of $f_c$ and the reduced $Q_i$ observed at higher temperatures are due to a decrease in Cooper pair density and an increase in the quasiparticle density in the MoRe film, respectively [18, 29].

The fitting results of $\alpha$ using the M-B theory are shown in Fig. 4c as a function of MoRe film thickness $t$. It is noted that $\alpha$ increases with decreasing film thickness. The corresponding $L_k$ values, shown in the inset, are inversely proportional to $t$ (see the dashed line) except for the thinnest film with $t = 15$ nm. The Fig. 4c-d including the inset figures display the data averaged over six resonators for each thickness. The inset of Fig. 4d displays $f_c$ as a function of $t$. Since the design of the SCPW resonators is the same for all samples studied in this work, the observed change of $f_c$ is due to an increase of $L_k$ with decreasing $t$. $L_k$ and $\alpha$ calculated from $f_c$ data are also displayed in the inset of Fig. 4c, which are consistent with the ones obtained from the fit using the M-B theory. It should be noted that $L_k$ becomes larger than $L_g \approx 432$ nH/m (see Table 1) for the resonators with $t < 30$ nm. The increased $L_k$ in thinner films induces an increase of $Z_0$ of the resonator circuit, as shown in Fig. 4d, which could be another cause of the suppression of $Q_i$ in thinner films in Fig. 4b.

Under the application of magnetic field, $B_\parallel$, parallel to the substrate, the progressive change of $S_{21}$ data is displayed in Fig. 5a for the MoRe resonator with $t = 60$ nm. The resonance dip is gradually broadened and the corresponding $f_c$ shifts to lower frequency with increasing $B_\parallel$, which is similar behavior to the temperature dependence of $|S_{21}|$ in Fig. 3c. After the $S_{21}$ data were converted into $S'_{21}$ data in Fig. 5b, the S-parameter fit was performed to extract $f_c$ and $Q_i$ values as a function of $B_\parallel$. The fractional $f_c$ shifts are shown in Fig. 5c as a function of $B_\parallel$ for the SCPW resonators with different $t$. All the resonator samples exhibit a negative shift of $f_c$, indicating an increase of $L_k$ with increasing $B_\parallel$ field. It has been known that the magnetic field increases the rate of Cooper-pair breaking, resulting in an increase of $\lambda_L$ and $L_k$ in NbTiN superconducting films [12, 32]. The resultant frequency shift can be approximated by $\delta f_{c,B}/f_{c,0} \approx -0.25(B_\parallel/B_{c2,\parallel})^2$, where $B_{c2}$ is the upper critical field [33]. Using the parabolic relation, we could extract $B_{c2,\parallel} = 3.9$ T for the MoRe film with $t = 89$ nm (see the solid line), which is consistent with previous work [34].

The magnetic field dependence of $Q_i$ is shown in Fig. 5d for the samples with different film thicknesses. When the magnetic field is applied parallel to the substrate, $Q_i$ decreases slowly up to the magnetic field threshold, $B_{th}$, and then drops sharply for $B_\parallel > B_{th}$ (see the arrow indicating $B_{th}$ for the sample with $t = 120$ nm). The threshold fields obtained in Fig. 5d are $B_{th} = 190, 54, 65, 930$, and $51$ mT for the samples with $t = 27, 41, 60, 89$, and $120$ nm, respectively. For the sample with $t = 15$ nm, it is hard to determine $B_{th}$ due to the lack of data points. Since there is no significant correlation found between $B_{th}$ and $t$, we infer that $B_{th}$ would be related to the possible formation of vortices in MoRe film due to a perpendicular magnetic field generated by an inaccurate alignment of the magnetic field direction with respect to the plane of the MoRe film.

It has been known that the lower critical magnetic field in a direction perpendicular to the surface of the superconducting thin film can be expressed as $B_{c1,\perp} = \Phi_0 \ln(\lambda_{eff}/\xi)/4\pi\lambda_{eff}^2$, where $\xi$ is the superconducting coherence length [35] and $\lambda_{eff} = \lambda_L^2/t$ is an effective penetration depth in thin film satisfying $t \ll \lambda_L$ [20]. We can then obtain $B_{c1,\perp} = 0.1, 0.2, 0.5, 1.2$, and $2.4$ mT and the probable misalignment angles, $\theta_{mis} = \sin^{-1}(B_{c1,\perp}/B_{th}) = 0.02, 0.2, 0.5, 0.07$, and $2.7$ degree, for the samples with $t = 27, 41, 60, 89$, and $120$ nm, respectively. The maximum density of vortices, which can be trapped by the antidot arrays formed in the ground plane, corresponds to the maximum trapping field, $B_{\perp,trap} = 14$ µT, which is much smaller than $B_{c1,\perp}$. Thus it is inferred that the vortices would be itinerant for $B_\parallel > B_{th}$, resulting in the excess loss in the resonator and a substantial reduction of $Q_i$ [11].

For a more quantitative comparison of the field-resilient characteristics of the MoRe resonators with different film thicknesses, we introduce the characteristic field $B_\parallel^*$ at which the $Q_i(B_\parallel^*)$ becomes half of the zero-field value, i.e. $Q_i(B_\parallel^*) = Q_i(0)/2$. The resultant $B_\parallel^*$, which were averaged over six different resonators for each film thickness, are displayed in Fig. 5e with the averaged $Q_i(B_\parallel = 0)$. It is noted that the maximum $B_\parallel^* = 0.85$ T is obtained from the sample with $t = 89$ nm, while the maximum $Q_i(0) = 2.1 \times 10^4$ is obtained from the resonator with $t = 41$ nm. Moreover, the optimal performance satisfying both criteria of $B_\parallel^* > 0.1$ T and $Q_i > 10^4$, as depicted by dashed lines in Fig. 5e, is achieved by the sample with $t = 27$ nm, which would be suitable for applications in spin qubits [14, 19] and a spin ensemble memory devices [8, 23].

For superconducting films whose thickness is thinner than $\lambda_L$, the in-plane lower critical field can be expressed by $B_{c1,\parallel} = (2\Phi_0/\pi t^2)\ln(t/1.07\xi)$ [21]. Calculation results of $B_{c1,\parallel}$ are shown in Fig. 5f for a perfect alignment of $B_\parallel$, revealing an increase of $B_{c1,\parallel}$ with decreasing film thickness. When the $B$ field is not perfectly aligned parallel to the surface of the superconducting film, it is necessary to consider the effect of $B_{c1,\perp}$ to determine $B_{th}$, which is the maximum $B_\parallel$ field for the usage of an SCPW resonator with a high $Q_i$ value. Figure 5f displays the calculation result of the nominal $B_\parallel$ field strength, $B_{vortex}$ (solid symbols), resulting $B_{c1,\perp}$ at each angle deviated from the parallel alignment, as a function of the film thickness. We note that $B_{th}$ would be determined by a smaller of $B_{c1,\parallel}$ and $B_{vortex}$ for the case with a small misalignment error. As a result, the highest value of $B_\parallel^*$ for a given misalignment angle would be obtained at the crossing point between $B_{c1,\parallel}$ and $B_{vortex}$ curves in Fig. 5f, as denoted by arrows.

**Conclusion**

In this study, we fabricated MoRe SCPW resonators with various film thicknesses and investigated their microwave transmission characteristics with varying temperature and external magnetic field. Temperature dependences of the $S_{21}$ spectrum are in good agreement with the Mattis–Bardeen theory, revealing that $L_k$ exceeds $L_g$ and the impedance mismatch occurs for the resonator sample with $t < 30$ nm. The negative shift of $f_c$ and the reduced $Q_i$ at higher temperatures are attributed to a decrease in Cooper pair density and an increase in the quasiparticle density. Applied with an external magnetic field parallel to the surface of MoRe film, the SCPW resonators show similar behavior of the negative shift of $f_c$ and the reduced $Q_i$. We found that the resonator with $t = 27$ nm seemed to be the best for a magnetic-field-resilient MoRe SCPW resonator, exhibiting $Q_i = 1.5 \times 10^4$ under $B_\parallel = 0.15$ T, which would be suitable for various applications in spin qubits, spin ensemble memory devices, and superconducting qubits made of graphene and CNTs.

**ACKNOWLEDGEMENTS**

This study was supported by the NRF of Korea through the Basic Science Research Program (2018R1A3B1052827, 2020M3E4A1106449, 2021M3H3A1037899, 2016R1A5A1008184) and the Ministry of Science and ICT under the ITRC program (IITP-2022-RS-2022-00164799).

**Figure caption**

**Figure 1.** (a) Resistivity vs. temperature curves and (b) residual resistivity, $\rho_{res}$, and $T_c$ of MoRe films with different film thicknesses. $\rho_{res}$ was measured at $T = 11$ K and $T_c$ was determined using the 10% criterion. The error bar means the transition width between 90% and 10% of the normalized resistance.

**Figure 2.** Scanning electron microscopy images of MoRe SCPW resonators. (a) Overall view of the sample. Six $\lambda/4$ resonators are coupled to a central feedline. Magnified images of (b) a single SCPW resonator, (c) an elbow-type end of the resonator shorted to the ground plane, (d) the opposite end of the resonator with an open-circuited stub, (e) the central feed line and the square-hole arrays for trapping vortices in the ground plane.

**Figure 3.** (a) Six resonance dips of 60-nm-thick MoRe resonator at $T = 1.8$ K. (b) $|S'_{21}|$ data (open dots) and the $S$-parameter fitting result (solid line) for R3 resonator in (a). We obtained $f_c = 3.61855$ GHz, $Q_i = 1.7 \times 10^4$ and $Q_c = 1.8 \times 10^3$ from the fit. Temperature dependences of (c) the R3 resonance dip and (d) $|S'_{21}|$ vs. $f - f_c$ plots (open dots) with the fitting results (solid lines).

**Figure 4.** Temperature dependences of (a) the fractional $f_c$ shifts and (b) the $Q_i$ values for the MoRe resonators with different film thicknesses. Symbols are experimental data, and the solid lines are from the theoretical fits to the M–B theory (see text). Film thickness dependences of (c) the kinetic inductance fraction $\alpha$ and (d) the characteristic impedance $Z_0$ at $T = 1.8$ K. $\alpha$ was obtained from the fit using the M–B theory (solid circles) and the $f_c$ averaged over the six resonators for the respective $t$, $<f_c>$. Insets: film thickness dependence of (c) $L_k$ and (d) $<f_c>$.

**Figure 5.** Magnetic field, $B_\parallel$, dependences of (a) the resonance dip for 60-nm-thick MoRe resonator, (b) $|S'_{21}|$ vs. $f - f_c$ plots (closed symbols) with the fitting results (solid lines), (c) the fractional $f_c$ shifts with the parabolic fit (see text) and (d) the $Q_i$ values for the MoRe resonators with different film thicknesses. (e) $B_\parallel^*$ vs. $Q_i(B_\parallel=0)$ plot. Symbols represent mean values obtained by averaging over the six resonators for the respective film thickness $t$. (f) Calculation results of $B_{c1,\parallel}$ and $B_{vortex}$ as a function of $t$. The misalignment angle, $\theta_{mis}$, responsible for $B_{vortex}$ is denoted in the legend.

**Table 1.** Physical parameters of MoRe SCPW resonators used in this work.

**Figure 1.**

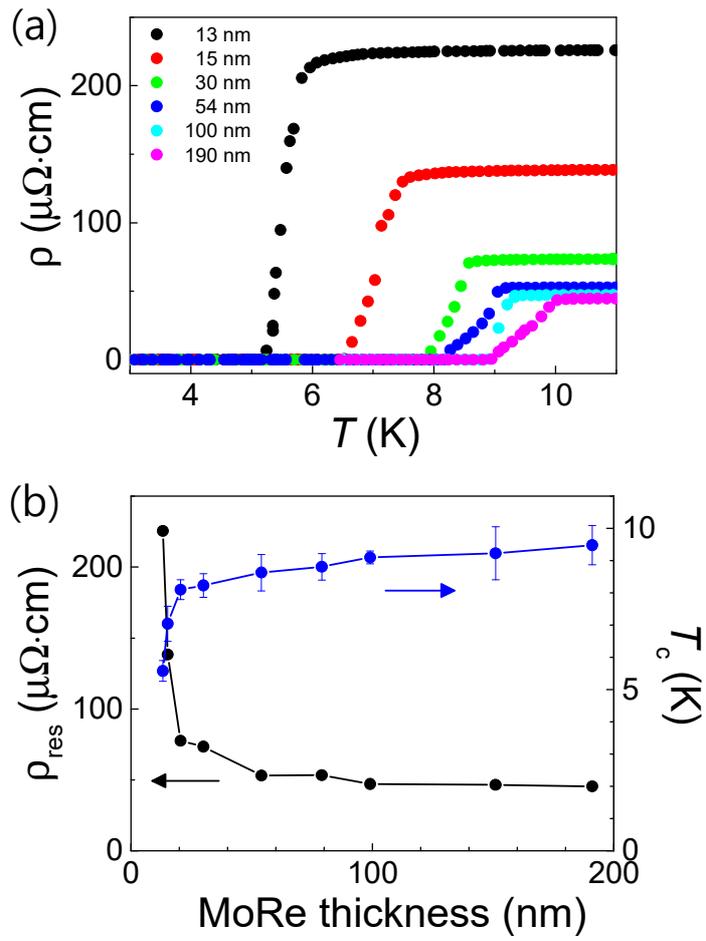

**Figure 2.**

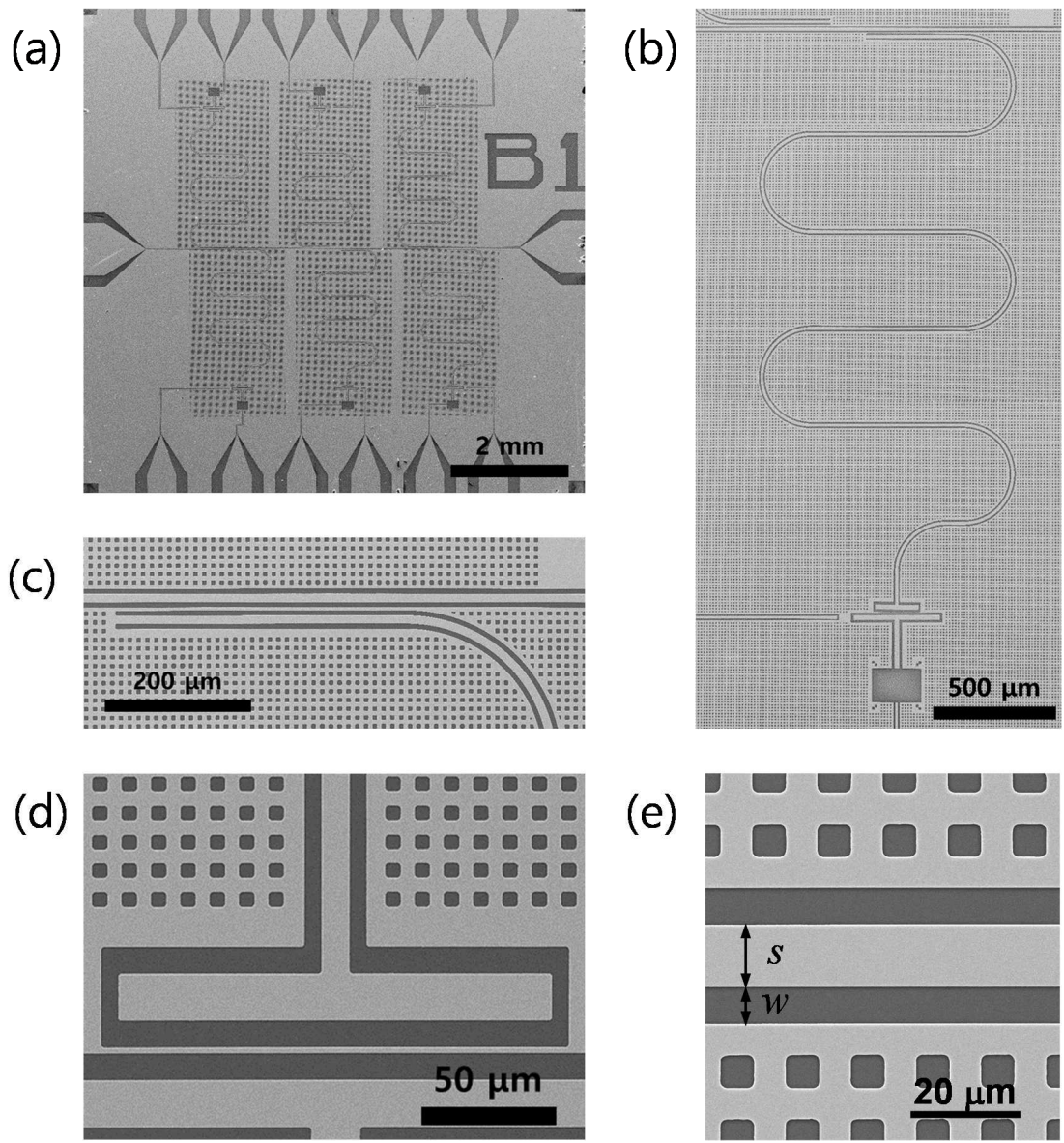

**Figure 3.**

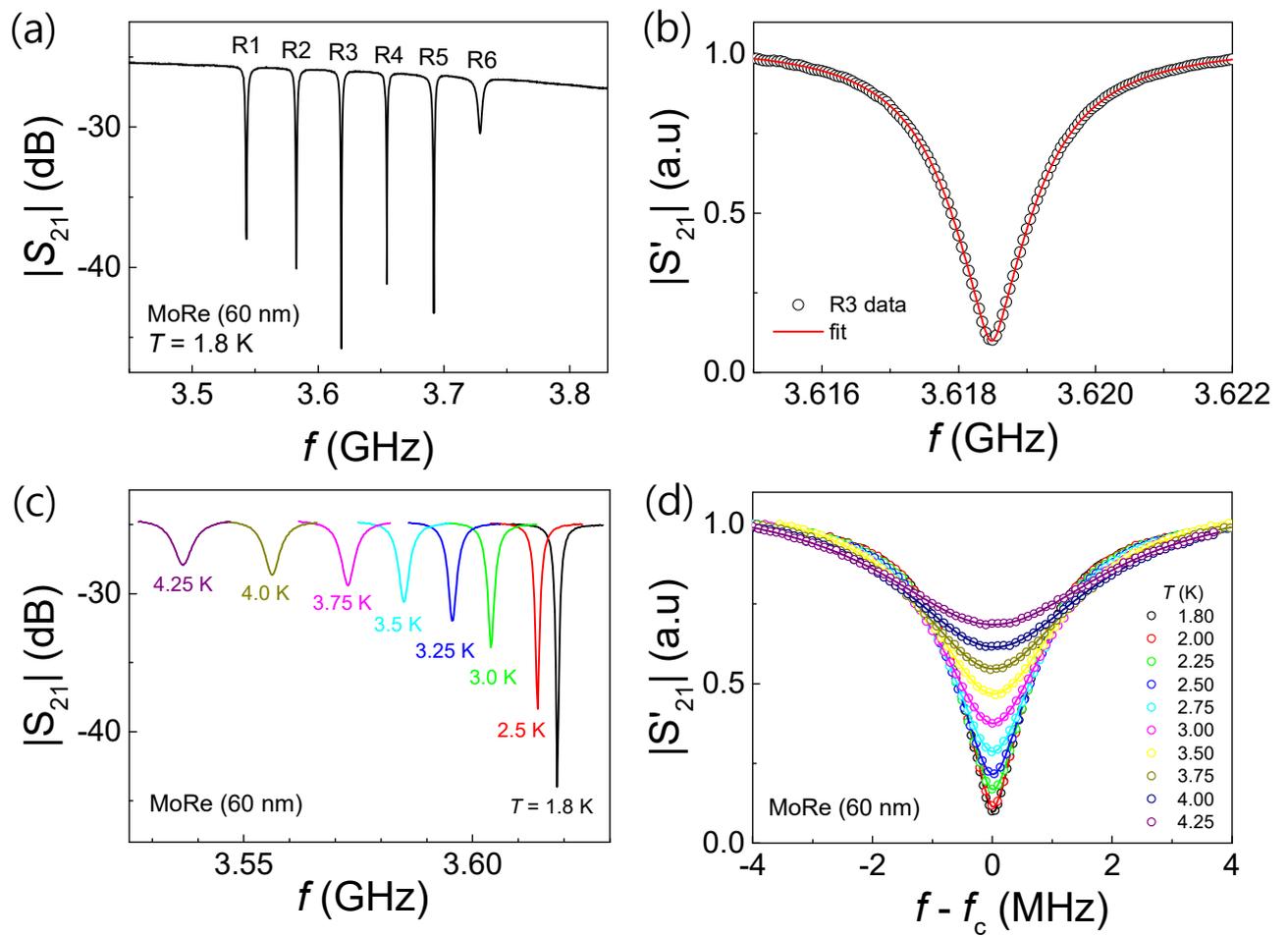

**Figure 4.**

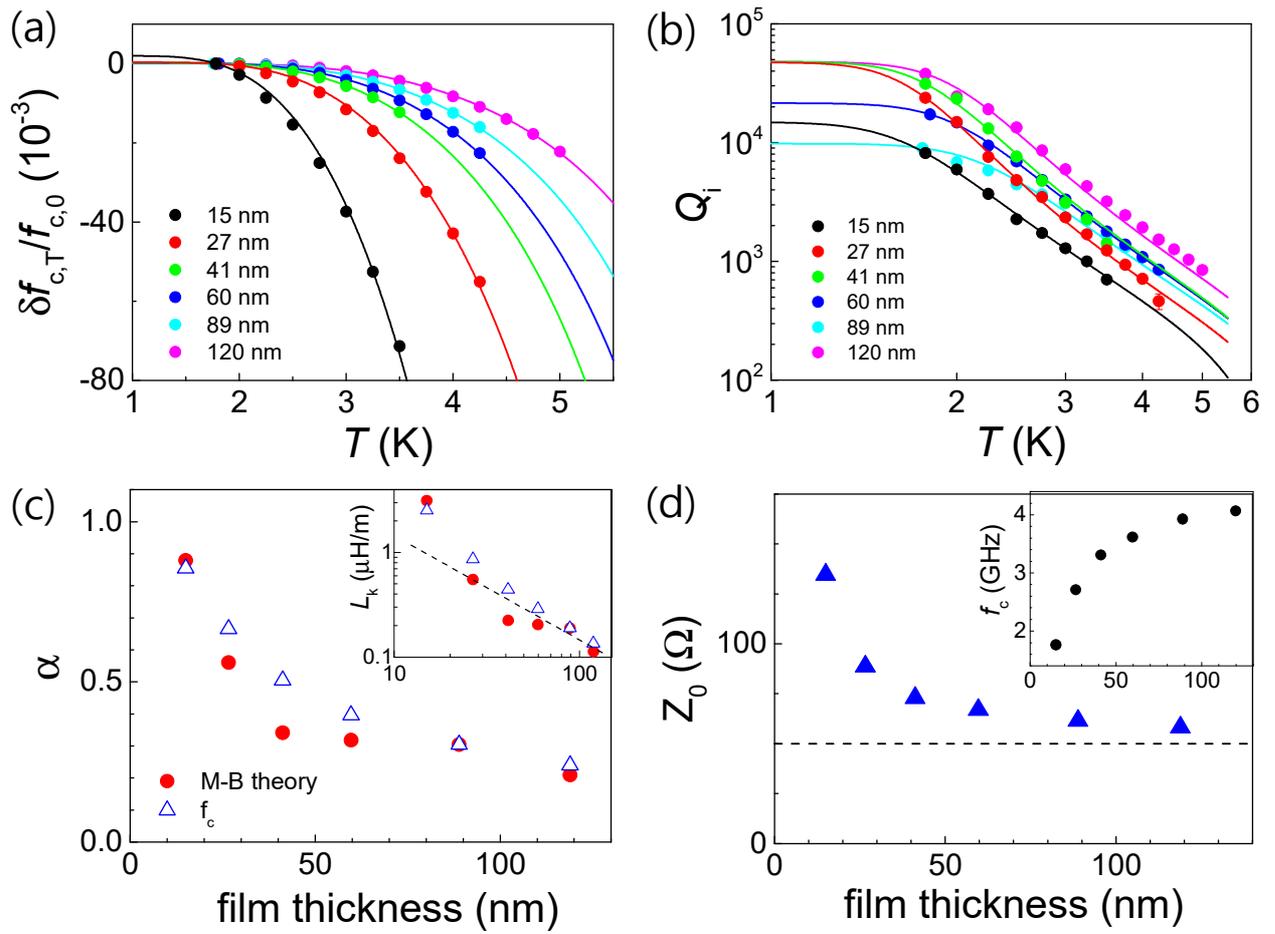

**Figure 5.**

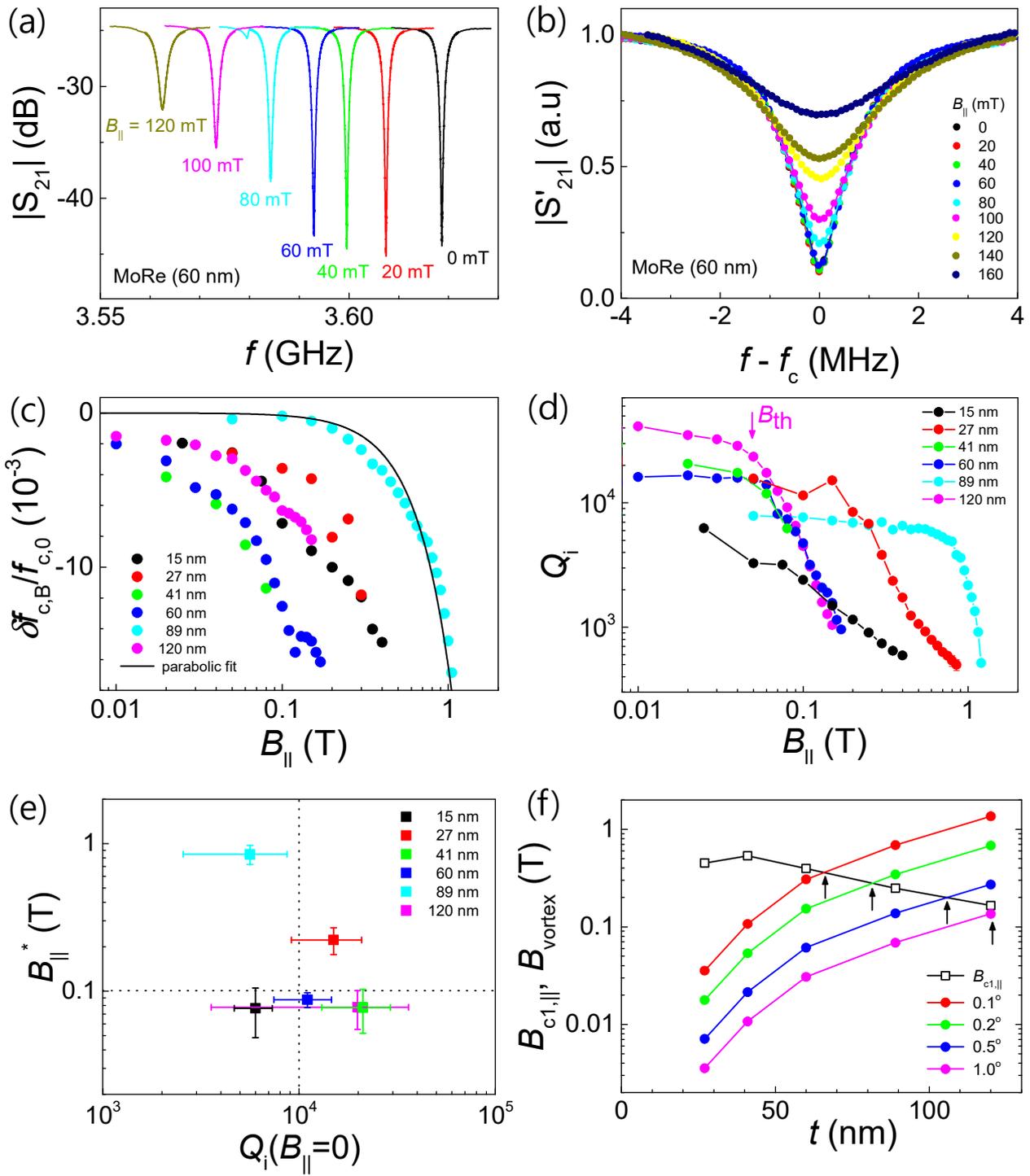

**Table 1.**

| No. | $t$ (nm) | $s$ (μm) | $w$ (μm) | $L_g$ (nH/m) | $C_g$ (pF/m) | $\rho_{res}$ (μΩ·cm) | $T_c$ (K) | $Z_0$ (Ω) |
|---|---|---|---|---|---|---|---|---|
| 1 | 120 | 11.9 | 6.6 | 428 | 167 | 47 | 9.2 | 58 |
| 2 | 89 | 11.7 | 6.7 | 432 | 165 | 50 | 9.0 | 61 |
| 3 | 60 | 11.6 | 7.0 | 438 | 163 | 53 | 8.8 | 67 |
| 4 | 41 | 11.8 | 6.8 | 432 | 165 | 64 | 8.5 | 73 |
| 5 | 27 | 11.7 | 6.8 | 433 | 165 | 75 | 8.3 | 89 |
| 6 | 15 | 11.8 | 6.6 | 429 | 166 | 147 | 6.9 | 135 |

References

[1] C.P. Wen, Coplanar waveguide: A surface strip transmission line suitable for nonreciprocal gyromagnetic device applications, IEEE Trans Microw Theory Tech, 17 (1969) 1087-1090.

[2] A. Megrant, C. Neill, R. Barends, B. Chiaro, Y. Chen, L. Feigl, J. Kelly, E. Lucero, M. Mariantoni, P.J.J. O'Malley, D. Sank, A. Vainsencher, J. Wenner, T.C. White, Y. Yin, J. Zhao, C.J. Palmstrøm, J.M. Martinis, A.N. Cleland, Planar superconducting resonators with internal quality factors above one million, Appl. Phys. Lett., 100 (2012) 113510.

[3] A. Wallraff, D.I. Schuster, A. Blais, L. Frunzio, R.-S. Huang, J. Majer, S. Kumar, S.M. Girvin, R.J. Schoelkopf, Strong coupling of a single photon to a superconducting qubit using circuit quantum electrodynamics, Nature, 431 (2004) 162-167.

[4] T.W. Larsen, K.D. Petersson, F. Kuemmeth, T.S. Jespersen, P. Krogstrup, J. Nygård, C.M. Marcus, Semiconductor-nanowire-based superconducting qubit, Phys. Rev. Lett., 115 (2015) 127001.

[5] G. de Lange, B. van Heck, A. Bruno, D.J. van Woerkom, A. Geresdi, S.R. Plissard, E.P.A.M. Bakkers, A.R. Akhmerov, L. DiCarlo, Realization of microwave quantum circuits using hybrid superconducting-semiconducting nanowire Josephson elements, Phys. Rev. Lett., 115 (2015) 127002.

[6] K.D. Petersson, L.W. McFaul, M.D. Schroer, M. Jung, J.M. Taylor, A.A. Houck, J.R. Petta, Circuit quantum electrodynamics with a spin qubit, Nature, 490 (2012) 380-383.

[7] F. Borjans, X.G. Croot, X. Mi, M.J. Gullans, J.R. Petta, Resonant microwave-mediated interactions between distant electron spins, Nature, 577 (2020) 195-198.

[8] V. Ranjan, G. de Lange, R. Schutjens, T. Debelhoir, J.P. Groen, D. Szombati, D.J. Thoen, T.M. Klapwijk, R. Hanson, L. DiCarlo, Probing dynamics of an electron-spin ensemble via a superconducting resonator, Phys. Rev. Lett., 110 (2013) 067004.

[9] D. Sabonis, O. Erlandsson, A. Kringhøj, B. Van Heck, T.W. Larsen, I. Petkovic, P. Krogstrup, K.D. Petersson, C.M. Marcus, Destructive Little-Parks Effect in a Full-Shell Nanowire-Based Transmon, Phys. Rev. Lett., 125 (2020) 156804.

[10] L. Frunzio, A. Wallraff, D. Schuster, J. Majer, R. Schoelkopf, Fabrication and characterization of superconducting circuit QED devices for quantum computation, IEEE Trans. Appl. Supercond., 15 (2005) 860-863.

[11] C. Song, T.W. Heitmann, M.P. DeFeo, K. Yu, R. McDermott, M. Neeley, J.M. Martinis, B.L. Plourde, Microwave response of vortices in superconducting thin films of Re and Al, Phys. Rev. B, 79 (2009) 174512.

[12] J.G. Kroll, F. Borsoi, K.L. van der Enden, W. Uilhoorn, D. de Jong, M. Quintero-Pérez, D.J. van Woerkom, A. Bruno, S.R. Plissard, D. Car, E.P.A.M. Bakkers, M.C. Cassidy, L.P. Kouwenhoven, Magnetic-field-resilient superconducting coplanar-waveguide resonators for hybrid circuit quantum electrodynamics experiments, Phys. Rev. Appl., 11 (2019) 064053.

[13] J.I.J. Wang, D. Rodan-Legrain, L. Bretheau, D.L. Campbell, B. Kannan, D. Kim, M. Kjaergaard, P. Krantz, G.O. Samach, F. Yan, J.L. Yoder, K. Watanabe, T. Taniguchi, T.P. Orlando, S. Gustavsson, P. Jarillo-Herrero, W.D. Oliver, Coherent control of a hybrid superconducting circuit made with graphene-based van der Waals heterostructures, Nat. Nanotechnol., 14 (2019) 120-125.

[14] E.A. Laird, F. Pei, L.P. Kouwenhoven, A valley–spin qubit in a carbon nanotube, Nat. Nanotechnol., 8 (2013) 565-568.

[15] V.E. Calado, S. Goswami, G. Nanda, M. Diez, A.R. Akhmerov, K. Watanabe, T. Taniguchi, T.M. Klapwijk, L.M.K. Vandersypen, Ballistic Josephson junctions in edge-contacted graphene, Nat. Nanotechnol., 10 (2015) 761-764.

[16] B.H. Schneider, S. Etaki, H.S.J. van der Zant, G.A. Steele, Coupling carbon nanotube mechanics to a superconducting circuit, Scientific Reports, 2 (2012) 599.

[17] F.E. Schmidt, M.D. Jenkins, K. Watanabe, T. Taniguchi, G.A. Steele, A ballistic graphene superconducting microwave circuit, Nat. Commun., 9 (2018) 1-7.

[18] V. Singh, B.H. Schneider, S.J. Bosman, E.P.J. Merkx, G.A. Steele, Molybdenum-rhenium alloy based high-Q superconducting microwave resonators, Appl. Phys. Lett., 105 (2014) 222601.

[19] L. Banszerus, K. Hecker, S. Möller, E. Icking, K. Watanabe, T. Taniguchi, C. Volk, C. Stampfer, Spin relaxation in a single-electron graphene quantum dot, Nat. Commun., 13 (2022) 3637.

[20] M. Tinkham, Introduction to superconductivity, Courier Corporation, 2004.

[21] G. Stejic, A. Gurevich, E. Kadyrov, D. Christen, R. Joynt, D.C. Larbalestier, Effect of geometry on the critical currents of thin films, Phys. Rev. B, 49 (1994) 1274-1288.

[22] N. Samkharadze, G. Zheng, N. Kalhor, D. Brousse, A. Sammak, U.C. Mendes, A. Blais, G. Scappucci, L.M.K. Vandersypen, Strong spin-photon coupling in silicon, Science, 359 (2018) 1123-1127.

[23] R. Amsüss, C. Koller, T. Nöbauer, S. Putz, S. Rotter, K. Sandner, S. Schneider, M. Schramböck, G. Steinhauser, H. Ritsch, J. Schmiedmayer, J. Majer, Cavity QED with Magnetically Coupled Collective Spin States, Phys. Rev. Lett., 107 (2011) 060502.

[24] B. Kim, M. Jung, J. Kim, J. Suh, Y.-J. Doh, Fabrication and characterization of superconducting coplanar waveguide resonator, Progress in Superconductivity and Cryogenics, 22 (2020) 10.

[25] G. Baek, B. Kim, S. Arif, Y.-J. Doh, Effect of capacitive coupling in superconducting coplanar waveguide resonator, Progress in Superconductivity and Cryogenics, 23 (2021) 6.

[26] D. Bothner, T. Gaber, M. Kemmler, D. Koelle, R. Kleiner, Improving the performance of superconducting microwave resonators in magnetic fields, Appl. Phys. Lett., 98 (2011) 102504.

[27] M. Göppl, A. Fragner, M. Baur, R. Bianchetti, S. Filipp, J.M. Fink, P.J. Leek, G. Puebla, L. Steffen, A. Wallraff, Coplanar waveguide resonators for circuit quantum electrodynamics, J. Appl. Phys., 104 (2008) 113904.

[28] K. Watanabe, K. Yoshida, T.A. Kohjiro, Kinetic inductance of superconducting coplanar waveguides, Jpn. J. Appl. Phys., 33 (1994) 5708.

[29] K.J.G. Götz, S. Blien, P.L. Stiller, O. Vavra, T. Mayer, T. Huber, T.N.G. Meier, M. Kronseder, C. Strunk, A.K. Hüttel, Co-sputtered MoRe thin films for carbon nanotube growth-compatible superconducting coplanar resonators, Nanotechnology, 27 (2016) 135202.

[30] D.C. Mattis, J. Bardeen, Theory of the anomalous skin effect in normal and superconducting metals, Phys. Rev., 111 (1958) 412.

[31] J. Gao, J. Zmuidzinas, A. Vayonakis, P. Day, B. Mazin, H. Leduc, Equivalence of the effects on the complex conductivity of superconductor due to temperature change and external pair breaking, J. Low. Temp. Phys., 151 (2008) 557-563.

[32] N. Samkharadze, A. Bruno, P. Scarlino, G. Zheng, D.P. DiVincenzo, L. DiCarlo, L.M.K. Vandersypen, High-kinetic-inductance superconducting nanowire resonators for circuit QED in a magnetic field, Phys. Rev. Appl., 5 (2016) 044004.

[33] K. Borisov, D. Rieger, P. Winkel, F. Henriques, F. Valenti, A. Ionita, M. Wessbecher, M. Spiecker, D. Gusenkova, I.M. Pop, W. Wernsdorfer, Superconducting granular aluminum resonators resilient to magnetic fields up to 1 Tesla, Appl. Phys. Lett., 117 (2020) 120502.

[34] S. Sundar, L.S.S. Chandra, V.K. Sharma, M.K. Chattopadhyay, S.B. Roy, Electrical transport and magnetic properties of superconducting Mo52Re48 alloy, AIP Conference Proceedings, 1512 (2013) 1092-1093.

[35] R. Gaudenzi, J.O. Island, J.d. Bruijckere, E. Burzurí, T.M. Klapwijk, H.S.J.v.d. Zant, Superconducting molybdenum-rhenium electrodes for single-molecule transport studies, Appl. Phys. Lett., 106 (2015) 222602.